\documentclass[ams,amsmath,amssymb,twocolumn,superscriptaddress,pra]{revtex4-2}
\usepackage{graphicx}
\usepackage{amsmath}
\usepackage{amssymb}
\usepackage{amsfonts}
\usepackage{color}
\usepackage{dsfont}
\usepackage{soul}
\usepackage{hyperref}
\usepackage{tikz}
\usepackage{lipsum}
\usepackage{float}
\usetikzlibrary{positioning,shapes}
\DeclareMathOperator{\sinc}{sinc}

\newcommand{\overbar}[1]{\mkern 1.5mu\overline{\mkern-1.5mu#1\mkern-1.5mu}\mkern 1.5mu}

\begin{document}
%%%%%%%%%%%%%%%%%%%%%%%%%%%%%%%%%%%%%%%%%%%%%%%%%%%%%%%%%%%%%%%%%%%%%%
%%%%%%%%%%%%%%%%%%%%%%%%%%%%%%%%%%%%%%%%%%%%%%%%%%%%%%%%%%%%%%%%%%%%%%
\title{Two-colour spectrally multimode integrated SU(1,1) interferometer}
\date{\today}

\author{Alessandro Ferreri}
\affiliation{Institute for Quantum Computing Analytics (PGI-12), Forschungszentrum J\"ulich, 52425 J\"ulich, Germany}
\affiliation{Department of Physics, Paderborn University,
Warburger Strasse 100, D-33098 Paderborn, Germany}
\author{Polina~R.~Sharapova}
\affiliation{Department of Physics, Paderborn University,
Warburger Strasse 100, D-33098 Paderborn, Germany}
	\begin{abstract}
Multimode integrated interferometers have great potential for both spectral engineering and metrological applications. However, material dispersion of integrated platforms constitutes an obstacle which limits the performance and precision of such interferometers. At the same time, two-colour non-linear interferometers present an important tool for metrological applications, when measurements in a certain frequency range are difficult. 
In this manuscript, we theoretically develop and investigate an integrated multimode two-colour  SU(1,1) interferometer that operates in a supersensitive mode. By ensuring a proper design of the integrated platform, we suppress dispersion and thereby significantly increase the visibility of the interference pattern. We demonstrate that such an interferometer overcomes the classical phase sensitivity limit for wide parametric gain ranges, when up to $3*10^4$ photons are generated.
\end{abstract}
\maketitle
\section{Introduction}
Quantum Metrology is the subject devoted to the fabrication and optimisation of high-precision interferometers. The interest on both quantum metrology and interferometry has been growing rapidly, and nowadays it can be acknowledged as one of the trend topics in quantum optics of the last decades \cite{PhysRevLett.96.010401, doi:10.1116/5.0007577, T_th_2014, giovannetti2011advances}. Surely, a boost in this direction was brought about by the first detection of gravitational waves, which provided a further precious tool to our comprehension of the universe \cite{PhysRevLett.116.061102}.

Conveniently, interferometers could be classified in two classes, linear interferometers and nonlinear or SU(1,1) interferometers, according to the presence of  passive or active components, respectively \cite{luo2021quantum}. In particular, the linear interferometer is characterized by passive elements, such as beam splitters, which maintain the total number of photons throughout the entire interference process; whereas the nonlinear interferometer consists of active elements, for example, optical parametric amplifiers (OPAs), triggering the creation of photons. This fundamental difference affects the strategy that has to be employed to achieve optimal phase sensitivity, i.e. the precision to trace the interference pattern \cite{Chekhova:16}. Indeed, on the one hand, to beat the classical shot noise limit (SNL), the linear interferometer requires the presence of squeezed or quantum-correlated input states \cite{doi:10.1126/science.1104149}, such as NOON state \cite{doi:10.1080/00107510802091298}, with a specific photon statistics \cite{slussarenko2017unconditional, DEMKOWICZDOBRZANSKI2015345}. On the other hand, due to the generation of the squeezed vacuum state by the OPAs, the SU(1,1) interferometer can overcome the SNL without any sort of seed, namely with the quantum vacuum as the input state \cite{PhysRevA.33.4033}, thereby potentially reaching the Heisenberg limit (HL), which is the limitation imposed by the time-energy uncertainty principle \cite{PhysRevA.55.2598}.

An excellent candidate for the role of photon source in various linear and nonlinear interferometric schemes  is the parametric down-conversion (PDC), which is one of the most commonly exploited nonlinear processes for the creation of squeezed and entangled photons \cite{PhysRevA.50.23, PhysRevA.50.5122, PhysRevA.56.1534}. 
Photons emitted via PDC, typically called signal and idler photons, are generated in broadband spatial and spectral modes, whose role in interferometry has already been proved both theoretically and experimentally \cite{PhysRevA.91.043816, Frascella_2019, PhysRevA.101.053843, PhysRevA.97.053827, Ferreri_2020, PhysRevLett.117.183601, PhysRevX.10.031063}. As an example, it has been observed that the possibility of controlling the number of spectral modes, and therefore the amount of spectral/ temporal correlations between signal and idler photons, can determine the behaviour of the quantum interference pattern in the four-photon Hong-Ou-Mandel scenarios \cite{PhysRevA.100.053829}.

It is therefore reasonable to believe that any specific realisation of the SU(1,1) interferometer based on realistic nonlinear photon sources should take the multimode structure of the emitted radiation into account. In this direction, in \cite{Ferreri2021spectrallymultimode} it has already been demonstrated how a proper state engineering \cite{paterova2020nonlinear} of the photon source allows one to reduce the dispersion within the interferometer and therefore improve both the visibility of the interference pattern and the accuracy in the phase scanning. 
In addition, proper spectral engineering of PDC sources allows the creation of correlated photons of different colours. During the last decade, the realization of such non-degenerate PDC sources with specific spectral features has become widespread \cite{PhysRevA.98.063844, dvernik2021azimuthal, Luo_2015}, while their application in interference systems such as Hong-Ou-Mandel \cite{PhysRevA.67.022301} and nonlinear Mach-Zehnder interferometers \cite{PhysRevA.101.053843} seems to be very promising.

Finally, the interest in integrated devices is recently growing due to both their small footprint and high efficiency \cite{tanzilli2012, caspani2017integrated, obrien2013}. Such technologies are already utilized in quantum interference scenarios \cite{Sharapova_2017, Ono:19}, in the realization of non-degenerate PDC sources suitable for single-photon heralding \cite{Krapick_2013}, as well as in the generation of polarization-entangled photons \cite{Herrmann:13}. 

In this manuscript, we theoretically develop a spectrally multimode non-degenerate integrated SU(1,1) interferometer, whose signal and idler photons are spectrally distinguishable and orthogonally polarized. As an example, we consider potassium titanyl phosphate (KTP) waveguides, however, the concept of such an interferometer can be applied to various integrated platforms. We demonstrate that the presented device works in the supersensitive mode beyond the classical limit SNL.

%Orthogonally polarized photons are generated by type-II PDC process based on periodic poled potassium titanyl phosphate (ppKTP) waveguides. The device can work beyond the SNL. 

The paper is organised as follow: in section II, we discuss the control of the photon generation process by means of the poling period to ensure the spectral distinguishability of signal and idler photons and present a theoretical model for describing integrated two-colour SU(1,1) interferomters based on such non-degenerate PDC processes.
%we take advantage of the possibility to tailor the periodic pole of the waveguide in order to control the momentum conservation of the photon generation process, and split the central frequency of both signal and idler photons in two different frequencies, such that the distance between their spectral intensity peaks is greater than the full width half maximum (FWHM). This procedure ensures the spectrally distinguishability of signal and idler photons. 
In section III, we show how dispersion within the two-colour SU(1,1) interferometer can be suppressed by properly designing an integrated platform, we analyze the phase sensitivity of such a two-colour interferometer, and compare the obtained results with the performance of its degenerate counterpart. Finally, we conclude with section IV and comment on the quantum features of the presented interferometer.

\section{Theoretical model}
\begin{figure}[H]
\centering
\includegraphics[width=1\linewidth]{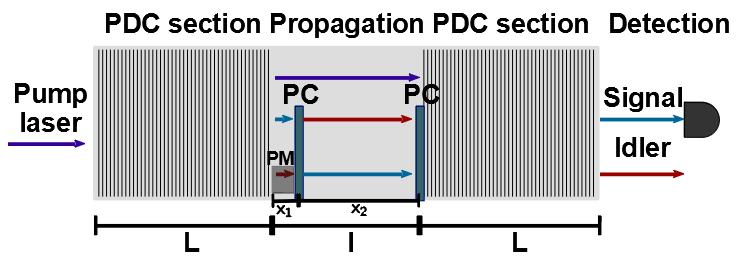}
\caption{Schematic model of the two-colour SU(1,1) interferometer. To produce spectrally non-degenerate and orthogonally polarized photons, the pump laser interacts with two PDC sections consisting of periodically poled KTP waveguides of length L. Different polarizations are depicted by the red and blue colours. The PDC sections are separated by the non-poled KTP section of length l. Two polarization converters, located at $x_1$ and $x_2$ positions, switch the polarizations of signal and idler photons in order to ensure a proper compensation of their group velocities when coming to the second PDC section. An additional phase for the idler photon is provided and controlled by the phase modulator PM. The signal photon is finally detected.}
\label{scheme}
\end{figure}

The integrated intereferometer is sketched in Fig.\ref{scheme}, and consists of two type-II PDC sources, in particular, two periodically poled potassium titanyl phosphate (ppKTP) waveguides, separated by a non-poled KTP section. The device is pumped by a continuous wave (CW) laser.
 The Hamiltonian describing the interferometer is given by
\cite{PhysRevA.97.053827, Ferreri2021spectrallymultimode}:
\begin{equation}
\hat H=i \hbar \Gamma \int d\omega_s d\omega_i F(\omega_s,\omega_i)\hat a_s^\dagger \hat a_i^\dagger+h.c.,
\label{Ham1}
\end{equation}
where $\Gamma$ is the coupling constant containing the second order susceptibility $\chi^{(2)}$, the pump intensity $I$ and the length of the PDC sections $L$; $\omega_s$ and $\omega_i$ are the frequencies of signal and idler photons, respectively, while $ \hat a_{s,i}^\dagger $ are the creation operators. The function $F(\omega_s,\omega_i)$ is the joint spectral amplitude (JSA), whose explicit expression will be obtained in the next section. The JSA contains both the energy and the momentum conservation of the PDC process, which are  expressed by the following relations:
\begin{align}
\omega_p-\omega_{s0}-\omega_{i0}=&0, \label{ec} \\
k_o(\omega_p)-k_o(\omega_{s0})-k_e(\omega_{i0})+\frac{2\pi}{\Lambda}=&0, \label{mc}
\end{align}
where $\omega_p$, $\omega_{s0}$, $\omega_{i0}$ are the central frequencies of the pump laser, signal and idler photons, respectively, while $k_x$ are their wave vectors. The indices $o$ and $e$ correspond to the ordinary and extraordinary polarization. The constant $\Lambda$ presents the waveguide poling period, which is required to close the phasematching condition (fulfill the momentum conservation), thereby enabling the pair production.
In the following analysis, we neglect the time ordering effect; which is a fair approximation for SU (1,1) interferometers,  taking into account the effective narrowing of their spectra due to the nonlinear interference \cite{PhysRevLett.117.183601}.

The lack of degeneracy characterizing two-colour spectra, namely $\omega_{s0}\neq\omega_{i0}$, is achieved by properly setting the poling period $\Lambda$ of the PDC section. 
This concretely means that the degree of freedom given by the choice of the periodic pole allows us to determine the central frequencies of the signal and idler photons, and consequently the separation of their spectra. In order to realise the fully spectral distinguishability of signal and idler photons, we firstly estimate the full width at half maximum (FWHM) of their spectra $\Delta\omega$ in a case of the degenerate PDC source with the poling period $\Lambda_{deg}$ \cite{Ferreri2021spectrallymultimode}, note that the spectra of signal and idler photons are identical in this case due to the use of the CW laser. Changing the poling period allows signal and idler photons to be generated at different central frequencies with a proper spectral separation. To characterize the splitting of the spectra, we impose a spectral detuning $\delta\omega$, namely a spectral distance between the degenerate frequency $\omega_p/2$ and the central frequency of the signal (idler) photon, $\omega_{s0}=\omega_p/2+\delta\omega$ and $\omega_{i0}=\omega_p/2-\delta\omega$, such that $\delta\omega\gg\Delta\omega$.
%As a poling period we therefore provisionally picked out the value expected for the degenerate PDC, $\Lambda_{deg}$, so that the central frequency of both signal and idler photons is momentary $\omega_p/2$.
This condition ensures the complete spectral splitting of signal and idler photons. The spectral detuning $\delta\omega$ is identical for both photons in order to fulfill the energy conservation of the process.
Eventually we consider the spectral detunings which are much smaller than the degenerate frequency $\omega_p/2$, so that $\Delta\omega\ll\delta\omega\ll\omega_p/2$. Then, the new value of the poling period $\Lambda$, which allows the photon pair creation at frequencies $\omega_{s0}$ and $\omega_{i0}$, is calculated by means of the momentum conservation in Eq \eqref{mc}. 
In contrast to the single-colour SU(1,1) interferometer described in \cite{Ferreri2021spectrallymultimode}, the current device is characterized by two polarization converters (PC) which are required for the proper compensation of the group velocities of signal and idler photons. The presence of two polarization conversions along the non-poled section enables the preparation of only one periodic pole. Indeed, if we use only one PC, the lack of degeneracy would push us to utilize a second waveguide with a poling period $\Lambda_2$ for the second PDC section, different with respect to the poling period $\Lambda$ of the first PDC section, because of different frequencies of the signal and idler photons. The presence of the second PC solves the problem, however, this PC must be properly arranged with respect to the first PC, whose exact position will be calculated in the next section. For the sake of simplicity, we can assume the second PC to be located in front of the second PDC section. 
The function describing the periodic pole is given by:
\begin{equation}
g(z)=\begin{cases}square_\Lambda(z) & 0<z<L   \\1 & L<z<l+L \\ square_\Lambda(z-(L+l)) & L+l<z<2L+l\end{cases}, 
\label{grate}
\end{equation}
where the periodic pole of the second PDC is not in phase with the periodic pole of the first one. This fact results in the generation of a global phase in the joint spectral amplitude (JSA).

In a general form, the phasematching function of the whole nonlinear interferometer of length $2L+l$ including several poled regions can be given by the expression \cite{klyshko1993ramsey, klyshko1994parametric, Santandrea_2019, Helmfrid:93}:
\begin{equation}
\Phi(\omega_s,\omega_i)= \int_{0}^{2L+l} dz\, g(z) e^{i\int_{0}^{z}d\xi \Delta k (\xi)},
\end{equation}
where $\Delta k (\xi)$ is the phasematching at the position $(\xi)$. Using Eq.\eqref{grate} and taking into account only the first-order terms in the Fourier decomposition of $g(z)$, the phasematching function describing the whole interferometer  presented in Fig.\ref{scheme} with  two periodically poled regions given by Eq.\eqref{grate} becomes: 
% \begin{equation}
%F(\omega_s,\omega_i)=\int_{0}^{L} dz\, e^{i\frac{2\pi z}{\Lambda}}e^{i\int_{0}^{z}d\xi \Delta k}+\int_{L+l}^{2L+l}dz\,e^{i\frac{2\pi z}{\Lambda}} e^{i\int_{0}^{z}d\xi\Delta k}
%\label{ff}
%\end{equation}

 \begin{align}
\Phi(\omega_s,\omega_i)=&\int_{0}^{L} dz\, e^{i\frac{2\pi z}{\Lambda}}e^{i\int_{0}^{z}d\xi \Delta k (\xi)}
\nonumber
\\
&+\int_{L+l}^{2L+l}dz\,e^{i\frac{2\pi (z-L-l)}{\Lambda}} e^{i\int_{0}^{z}d\xi\Delta k (\xi)}, 
\label{ff}
\end{align}
where the phasematching $\Delta k (\xi)$ takes the following expressions in different regions of the interferometer:
\begin{widetext}
 \begin{equation}
\Delta k (\xi)=\begin{cases} \Delta k= k_o(\omega_p)-k_o(\omega_s)-k_e(\omega_i) & 0<z<L   \\
\Delta k'= k_o(\omega_p)-k_o(\omega_s)-k_e'(\omega_i) & L<z<L+x_1   \\
\overbar{\Delta k}= k_o(\omega_p)-k_e(\omega_s)-k_o(\omega_i) & L+x_1<z<L+x_1+x_2 = L+l \\ \Delta k= k_o(\omega_p)-k_o(\omega_s)-k_e(\omega_i)& L+l<z<2L+l,\end{cases}
\label{Deltak}
\end{equation}
\end{widetext}
where  $k_e'(\omega_i)=k_e(\omega_i)+\delta k_e (\omega_i)$ is the modified wavevector of the idler photon in the phase modulation region. Such a change of the wavevector can be performed by applying external voltage. Hence, introducing an average phase imparted by external voltage $\varphi=\delta k_e (\omega_{i0})x_1$ with respect to the central frequency of the idler photon $\omega_{i0}$, one can define the difference of optical paths in the phase modulation region as $\Delta k' x_1 = \Delta k x_1 + \delta k_e (\omega_i) x_1 = \Delta k x_1 + \varphi \frac{\delta k_e (\omega_{i})}{\delta k_e (\omega_{i0})} \approx \Delta k x_1 + \varphi$, since the relation $\frac{\delta k_e (\omega_{i})}{\delta k_e (\omega_{i0})} \approx 1$ holds in the vicinity of $\omega_{i0}$.  

We now evaluate two integrals in Eq.\eqref{ff}. The first integral provides the JSA of the first PDC section:
\begin{align}
\int_{0}^{L} dz\, e^{i\frac{2\pi z}{\Lambda}}e^{i\int_{0}^{z}d\xi \Delta k (\xi)}=&\int_{0}^{L} dz\, e^{i\frac{2\pi z}{\Lambda}}e^{\Delta k z }\nonumber\\
= & L \sinc\left[\frac{\Delta\beta L}{2 }\right]e^{i\frac{\Delta \beta L}{2}},
\label{crystal1}
\end{align}
where $\Delta \beta=\Delta k +\frac{2\pi}{\Lambda}$.
%where $\Delta \beta=k_o(\omega_p)-k_e(\omega_{s})-k_o(\omega_{i})+\frac{2\pi}{\Lambda}$. 

\begin{figure}[H]
\centering
\includegraphics[width=1\linewidth]{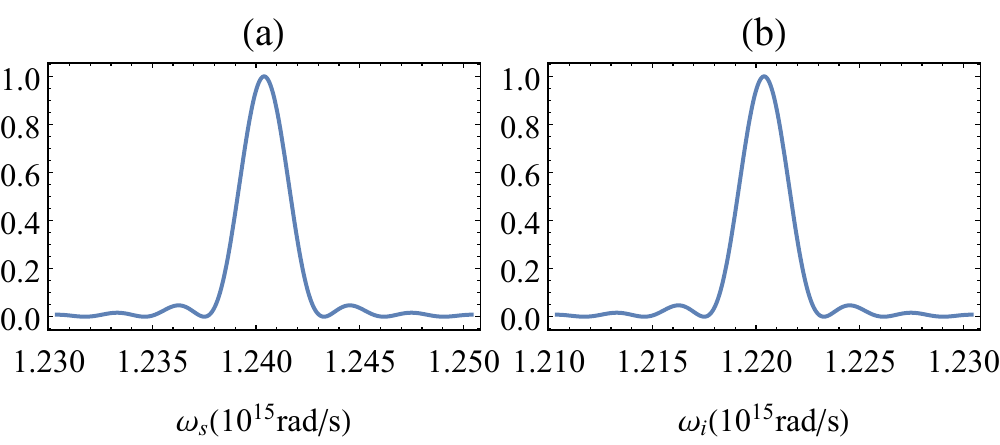}
\caption{Normalized spectra of (a) signal and (b) idler photons. The FWHM of both is $\Delta\omega$= 2*$10^{12}$ rad/s and the spectral detuning of both is $\delta\omega$=10*$10^{12}$ rad/s. The following parameters were used for calculations: L = 8 mm, $x_1\simeq 1.038$ mm, $x_2\simeq 8.962$ mm, $\Lambda=133\mu m$, pump wavelength $\lambda_p$ = 766 nm, $\phi=\varphi+\Phi=0$ (see Eq. \eqref{Phi}  for more details). }
\label{spectraf}
\end{figure}

The second integral can be evaluated as follows:

\begin{widetext}
\begin{align}
\int_{L+l}^{2L+l}dz\,e^{i\frac{2\pi (z-L-l)}{\Lambda}} e^{i\int_{0}^{z}d\xi\Delta k (\xi)}=& \int_{L+l}^{2L+l}dz\,e^{i\frac{2\pi (z-L-l)}{\Lambda}} e^{i\Delta k L}e^{i\Delta k' x_1}e^{i\overbar{\Delta k}x_2} e^{i\Delta k(z-x_1-x_2-L)} \nonumber\\
=& L \sinc\left[\frac{\Delta\beta L}{2 }\right]e^{i\frac{\Delta \beta L}{2}} e^{i\frac{2\pi (-L-l)}{\Lambda}} e^{i(\Delta k'-\Delta k) x_1}e^{i(\overbar{\Delta k}-\Delta k) x_2} e^{i\Delta \beta (L+l)}.
\label{crystal2}
\end{align}
\end{widetext}
Using the definition of the averaged phase introduced above and collecting Eqs.\eqref{ff}, \eqref{crystal1} and \eqref{crystal2}, one can write the expression for the joint spectral amplitude, which under the condition of the CW pump has the form: 

\begin{align}
F(\omega_s,\omega_i)=& C \delta(\omega_p-\omega_s-\omega_i)\sinc\bigg[\frac{ \Delta \beta L}{2}\bigg]\nonumber\\
&\times e^{i\frac{\Delta \beta L}{2}}\bigg[1+ e^{i\varphi}e^{i \overbar{\Delta k} x_2} e^{i\Delta k (L+x_1)}\bigg],
\label{JSAcomp}
\end{align}
where the $\delta$-function indicates the CW pump, while $C$ is the normalization constant. We notice that the length of the PDC source within the sinc-function fashions the width of the joint spectral intensity (JSI) and therefore the FWHM of the spectra of both signal and idler photons. A pondered choice of L therefore plays a crucial role in the realisation of the spectral distinguishability of signal and idler photons. The spectra of the signal (idler) photons can be calculated as follows 
\begin{align}
I(\omega_{s (i)})=\int d\omega_{i (s)} |F(\omega_s,\omega_i)|^2,
\label{spectra}
\end{align}
and are presented in Fig.\ref{spectraf} for the length L=8mm and the poling period $\Lambda=133\mu m$. 

%\begin{equation}\begin{split}
%=e^{i\Delta k(L+x_1)}e^{i\overbar{\Delta k}x_2} e^{i\Delta k(-x_1-x_2-L)} \int_{L+l}^{2L+l}dz\,e^{i \Delta \beta z}e^{i\frac{2\pi}{\Lambda}(-L-l)}\\
%=e^{i\Delta k(-x_2)}e^{i\overbar{\Delta k}x_2}e^{i\frac{2\pi}{\Lambda}(-L-l)}e^{i\Delta \beta (L+l)}   \sinc\left[\frac{\Delta\beta L}{2 }\right]e^{i\frac{\Delta \beta L}{2}}\\
%=e^{i\Delta k(-x_2)}e^{i\overbar{\Delta k}x_2}e^{i\Delta k (L+l)}   \sinc\left[\frac{\Delta\beta L}{2 }\right]e^{i\frac{\Delta \beta L}{2}}\\
%=e^{i\Delta k(L+x_1)}e^{i\overbar{\Delta k}x_2} \sinc\left[\frac{\Delta\beta L}{2 }\right]e^{i\frac{\Delta \beta L}{2}},
%\end{split}\end{equation}

%By resuming, the JSA describing the process is: 
%\begin{equation}\begin{split}
%F(\omega_s,\omega_i)=\frac{C}{2}\delta(\omega_p-\omega_s-\omega_i)\sinc\bigg[\frac{ \Delta \beta L}{2}\bigg]e^{i\Delta \beta L/2}\bigg[1+e^{i\Delta \beta L+i\Delta \beta' x_1+i\overbar{\Delta \beta}x_2}\bigg],
%\label{JSAcomp}
%\end{split}\end{equation}

\section{Dispersion suppression and phase sensitivity}
The regime of destructive interference of SU(1,1) interferometers is very important for reducing noise and improving phase sensitivity \cite{Chekhova:16}. Analyzing a degenerate multimode SU(1,1) interferometer, we have already found that the shape of the JSI and the number of generated photons at different imparted phase $\varphi$ are strongly conditioned by the arrangement of various components within the integrated device \cite{Ferreri2021spectrallymultimode}. 
In this section, to improve the phase sensitivity of the non-degenerate two-colour SU(1,1) interferometer, we elaborate a strategy for reducing the number of generated photons in the destructive interference regime by properly placing polarization converters.

We start by noticing that according to the last line of Eq.\eqref{JSAcomp}, the JSA is modulated by a cosine function, whose argument depends on $\varphi$. However, the two exponential terms in the end of the second line explicitly depend on the frequencies of both signal and idler photons; this causes by dispersion that prevents the suppression of photon generation in the destructive interference regime. Nevertheless, the use of proper positions of polarization converters $x_1$ and $x_2$ allows us to compensate for the first-order dispersion terms and thereby reduce the arguments of such exponential factors to a mere additional global phase.

%Since we assume the length of the PDC waveguides to be fixed, the only controllable terms in these exponential factors are $x_1$ and $x_2$. This means concretely that we have to find the proper positions of the polarization converters. 

To find such positions, we apply the Taylor expansion to the wavevectors of the signal and idler photons around their central frequencies $\omega_{s0}$ and $\omega_{i0}$, respectively. In the zeroth order we have
\begin{equation}\label{zerorder}
\Delta k^{(0)} (L+x_1)+\overbar{\Delta k}^{(0)}x_2 = -\frac{2\pi}{\Lambda} (L+x_1)+K x_2 \equiv \Phi, 
\end{equation}
where $K = k_o(\omega_p)-k_e(\omega_{s0})-k_o(\omega_{i0})$. One can observe that the zero-order term results in a global phase $\Phi$. In the following consideration, we will shift the value of the phase $\varphi$ by the global phase $\Phi$, in order to have the maximum of JSA Eq.\eqref{JSAcomp} for the phase $\phi=\varphi+\Phi = 0$.
%\textcolor{blue}{In order to start the modulation of the JSA from its maximum value, we shift the zero of the global phase $\phi$ equal to the opposite of the right side of Eq.\eqref{zerorder}}.

In the first order, one can write the following system of equations:
\begin{eqnarray}
\Delta k^{(1)} (L+x_1)+\overbar{\Delta k}^{(1)}x_2=0,
\label{1order}\\
x_1+x_2=l.
\label{1order1}
\end{eqnarray}

%where we momentary switched off the internal phase $\phi$. 

%Since both $\Delta k$ and $\overbar{\Delta k}$ depend on the frequencies of signal and idler photons, the first equation cannot be solved for all orders. However, we will solve it up to the first order. 

By exploiting the presence of strong frequency correlations between signal and idler photons caused by the use of CW pump laser, $\omega_i=\omega_p-\omega_s$, Eq.\eqref{1order} becomes:
\begin{equation}
\left(-\frac{\Omega}{v_{o,b}}+\frac{\Omega}{v_{e,r}}\right)(L+x_1)+\left(-\frac{\Omega}{v_{e,b}}+\frac{\Omega}{v_{o,r}}\right)x_2=0,
\label{1order2}
\end{equation}
where $\Omega=\omega_s-\omega_p/2-\delta\omega$ is the frequency detuning, and $v=(\partial k/\partial \omega)^{-1}$ are the group velocities calculated with respect to the central frequencies of photons. Here we determine the central frequencies of photons by the indices $r$ and $b$, namely, $r$ corresponds to the "red" idler photon $\omega_r \equiv \omega_{i0}=\omega_p/2-\delta\omega$,  while $b$ - to the "blue" signal photon $\omega_b \equiv \omega_{s0}=\omega_p/2+\delta\omega$. 
%This equation does not depend on signal frequency:
%\begin{equation}
%\left(-\frac{1}{v_{o,b}}+\frac{1}{v_{e,r}}\right)(L+x_1)+\left(-\frac{1}{v_{e,b}}+\frac{1}{v_{o,r}}\right)x_2=0.
%\end{equation}
The solution of the coupled equations \eqref{1order1},\eqref{1order2} allows us to find the proper positions of both polarization converters in order to suppress the first-order dispersion terms and obtain quasi-perfect interference at the output of the two-colour SU(1,1) interferometer:
\begin{align}
x_1=&\frac{ v_{e,r}v_{o,b}(v_{e,b}-v_{o,r})l+ v_{e,b}v_{o,r}(v_{o,b}-v_{e,r})L}{v_{e,b}v_{e,r}v_{o,b}+v_{e,b}v_{e,r}v_{o,r}-v_{e,b}v_{o,b}v_{o,r}-v_{e,r}v_{o,b}v_{o,r}},\\
x_2=&\frac{v_{o,r}v_{e,b}(v_{e,r}-v_{o,b})(l+ L)}{v_{e,b}v_{e,r}v_{o,b}+v_{e,b}v_{e,r}v_{o,r}-v_{e,b}v_{o,b}v_{o,r}-v_{e,r}v_{o,b}v_{o,r}}.
\end{align}

\begin{figure*}
	\centering
	\includegraphics[width=1\linewidth]{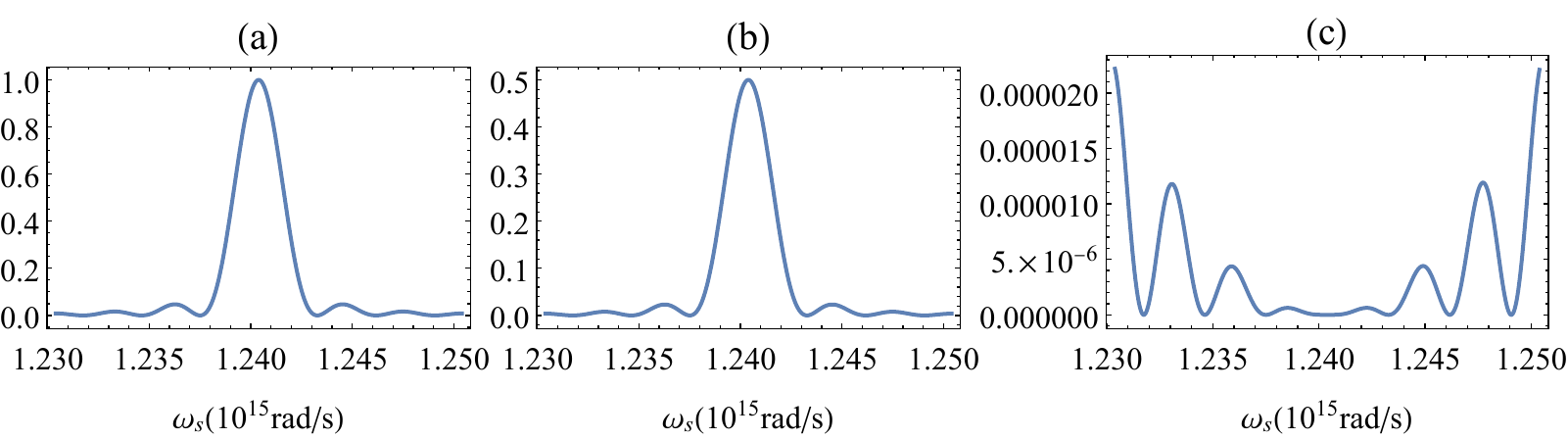}
	\caption{Normalized spectra of signal photons at (a) $\phi=0$, (b) $\phi=\pi/2$, and (c) $\phi=\pi$. The normalization is performed with respect to the maximum intensity in the constructive interference case, when $\phi=0$. The following parameters are chosen and fixed for all further calculations: $\lambda_p$ = 766 nm, L = 8 mm, $x_1\simeq 1.038$ mm, $x_2\simeq 8.962$ mm, $\Lambda=133\mu m$. The choice of $\delta\omega=10^{13}$rad/s ensures the fully spectral distinguishability of signal and idler photons.}
	\label{noseeding}
\end{figure*}

Fig.\ref{noseeding} presents the signal spectra normalized to the maximum intensity for three different phases corresponding to the regimes of constructive $\phi=0$ (Fig.\ref{noseeding} a) and destructive $\phi=\pi$ (Fig.\ref{noseeding} c) interference, as well as to the intermediate regime with the phase $\phi=\pi/2$ (Fig.\ref{noseeding} b). 
 As it can be observed in Fig.\ref{noseeding}c, the intensity of the signal photon in the destructive interference case is drastically reduced with respect to the constructive interference regime, although a noise at large frequency detunings $\lvert\Omega\rvert>0$ is present. The presence of this noise is caused by the fact that our technique for compensating the group velocities of signal and idler photons succeeds in the proximity of the central frequencies (namely where the main peaks of the intensity spectra are expected), whereas it fails for frequencies far from $\omega_{s0}$ and $\omega_{i0}$, resulting in the emergence of residual photons. The same issue has been already observed in the degenerate SU(1,1) interferometer \cite{Ferreri2021spectrallymultimode}. In both degenerate and two-colour frameworks, the presence of this residual radiation hinders the perfect interference, and therefore affects the precision of the interferometer, which can be estimated by means of the phase sensitivity:
\begin{equation}
\lvert\Delta\phi\rvert=\bigg\lvert\frac{\Delta N}{\partial \langle N \rangle/\partial\phi}\bigg\rvert,
\label{psdef}
\end{equation}
where $\Delta N$ is the standard deviation of the number of  photons, while $\langle N \rangle$ is the mean number of photons.

\begin{figure}[H]
	\centering
	\includegraphics[width=1\linewidth]{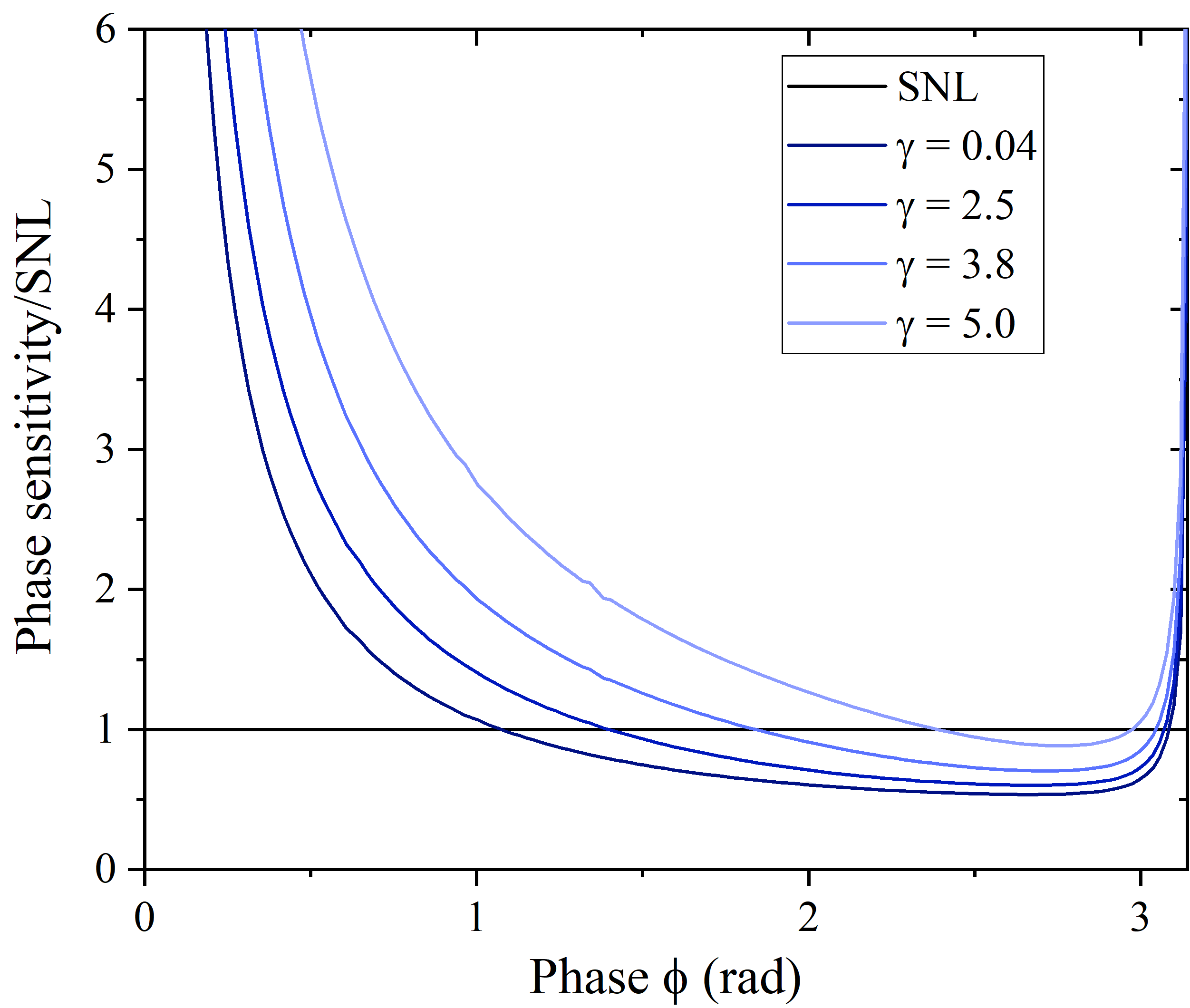}
	\caption{The phase sensitivity normalized to SNL versus phase at different gains. The SNL is shown by the black line. }
	\label{ps}
\end{figure}

Fig.\ref{ps} shows the trend of the phase sensitivity calculated with respect to the signal photons and normalized to the shot noise limit (SNL) when varying the internal phase of the interferometer. The shot noise limit, that determines the classical phase sensitivity bound, is calculated relative to the number of photons inside the interferometer $\langle N_{in} \rangle$:
\begin{equation}
\lvert\Delta\phi_{SNL}\rvert=\frac{1}{\sqrt{ \langle N_{in} \rangle}}.
\label{SNL}
\end{equation}
Since the phase sensitivity is a symmetrical function with respect to the point $\phi=\pi$, we only report the range $[0, \pi]$. As already experienced in \cite{Ferreri2021spectrallymultimode}, the normalized phase sensitivity diverges in the proximity of both $\phi=0$ and $\phi=\pi$. The presence of these divergences can be mathematically clarified by looking at Eq.\eqref{psdef}: At $\phi=0$ both the number of photons and its variance reach the maximum value, whereas the derivative is identically equal to zero; at $\phi=\pi$, the presence of residual photons prevents the variance of the photon number dropping to zero, whereas the derivative of $\langle N \rangle$ is identically equal to zero, being at the point of minimum intensity.

As can be seen from Fig.\ref{ps}, the presented interferometer can perform the phase scan in a supersensitive mode,  where the phase sensitivity overcomes the classical limit set by SNL. One can observe that the normalized phase sensitivity dramatically depends on the parametric gain, which is defined by $\gamma=G(\phi=0)\sqrt{\lambda_1(\phi=0)}$, where $G(\phi=0)$ includes both the coupling constant of the PDC process and the normalization factor of the JSA, and $\lambda_1(\phi=0)$ is the first eigenvalue of the spectral Schmidt-mode decomposition of JSA \cite{PhysRevLett.84.5304}. Both the parameter G and the first Schmidt eigenvalue $\lambda_1$ are evaluated at the maximum of intensity when $\phi=0$ \cite{Ferreri2021spectrallymultimode}. We stress that by varying the gain opportunely, we can let the number of photons to vary from $\langle N(\phi=0)\rangle\approx0.12$ for $\gamma\simeq0.04$, to $\langle N(\phi=0)\rangle\approx10^9$ for $\gamma\simeq10.0$.

\footnotetext[1]{Since the primary interest of this work is to develop a multimode description of two-colour integrated SU(1,1) interferometers, we decided to avoid the consideration of losses and focus our investigation on the role of  dispersion in the phase estimation problem. Accounting for losses, whose influence on the phase sensitivity is already well-known in literature, would require a considerable extension of the existing theory.}

\begin{figure}[H]
	\centering
	\includegraphics[width=1\linewidth]{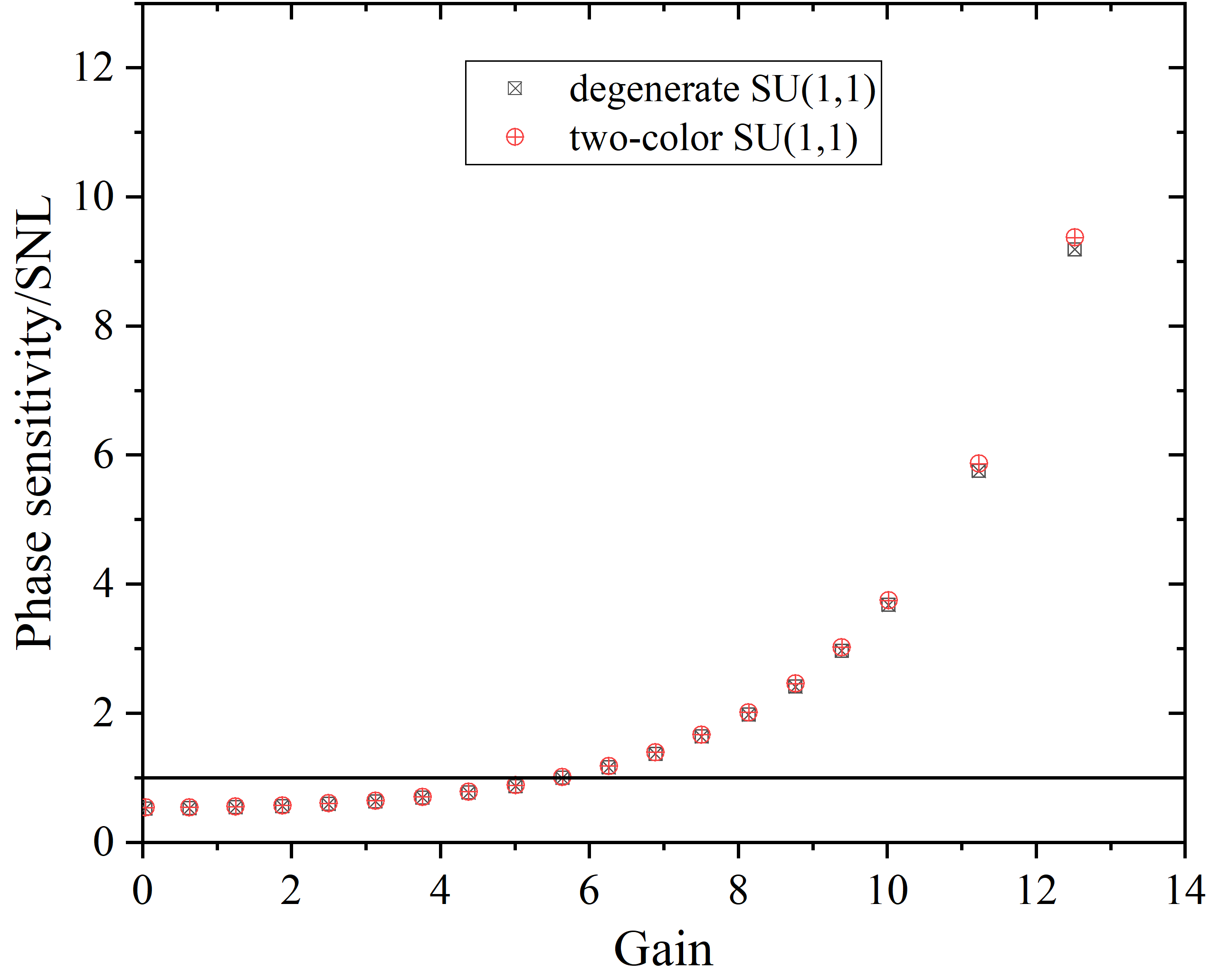}
	\caption{The minimum values of the normalized phase sensitivity presented in Fig.\ref{ps} versus gain $\gamma$ in comparison with the degenerate case reported in \cite{Ferreri2021spectrallymultimode}. The higher the gain is, the faster the phase sensitivity grows. The SNL is plotted by the black line.}
	\label{psg}
\end{figure}

The dependence of the normalized phase sensitivity on the gain is underlined in Fig.\ref{psg}. It can be seen that the two-colour SU (1,1) interferometer operates in the supersensitive regime for the same gain region as the degenerate (single-colour) SU (1,1) interferometer \cite{Ferreri2021spectrallymultimode}. Moreover, the phase sensitivity of both interferometers behaves in the same way, and its degradation is observed with an increase in the gain.
 We attribute the worsening of the performance at higher gains to the high-order dispersion terms, which cause the emergence of the central peak at $\phi=\pi$ in Fig. \ref{ps}, and prevent the improvement of the phase sensitivity with increasing gain, in contrast to the single plane-wave mode scenarios \cite{ manceau2017improving, PhysRevA.95.063843, Li_2014, Plick_2010}, where the worsening of the phase sensitivity is associated only with the presence of internal losses  \cite{Li:18, PhysRevA.86.023844, doi:10.1063/1.4960585, PhysRevA.98.023803, Note1}. However, despite deterioration in phase sensitivity at high gains, we notice that the presence of a supersensitivity range is maintained at gain values of about 4, namely, when the interferometer produces about 3*$10^4$ photons.

\section{Conclusion}
In this paper we modelled a high-performance integrated SU(1,1) interfermeter whose output photons are characterized by frequency distinguishability. The spectral separation of signal and idler photons was achieved by the correct choice of the periodic pole of both PDC sources composing the interferometer. At the same time, a compensation of the group velocities of photons was preformed by using two polarization converters placed at the appropriate positions inside the interferometer, that results in a suppression of dispersion and, consequently, leads to an improvement in the accuracy of the interferometer. We demonstrated that the phase sensitivity of the presented multimode non-degenerate interferometer overcomes the classical limit and reaches the same magnitudes  as in the case of a degenerate SU(1,1) interferometer.

Our results show that the possibility of fully controlling the intensity of the output radiation, the spectral characteristics and distinguishability of signal and idler photons offers a high degree of manipulation, and makes this device a useful photon source for a large variety of optical scenarios. Moreover, the discussed interferometer benefits from its reduced footprint due to the integrated design as well as from its outstanding performance in the phase scanning. The latter is confirmed by the fact that the phase sensitivity of the interferometer can overcome the shot noise limit, while the interferometer can still generate a considerable amount of photons. This specific feature of the presented interferometer is the result of a combination of two factors: the use of a CW laser and the presence of two polarization converters. By properly locating these  polarization converters, we were able to reduce dispersion and drastically decrease the number of photons in the destructive interference regime and thereby maximize the visibility of the interference pattern.

The presented two-colour SU(1,1) interferometer can be of wide interest for metrological applications where the measurement over one of the photon is challenging. The efficiency of this interferometer can be further tested by making use of different seeding strategies.
Furthermore, we leave the realisation of the high-performance two-colour SU(1,1) interferometer, whose emitted photons preserve both frequency and polarization entanglement, for future work.
Finally, this model will soon be upgrated  by accounting for internal losses, which typically further erodes the efficiency of interferometers.

\section{Acknowledgements}
We acknowledge the financial support of the Deutsche Forschungsgemeinschaft (DFG) via TRR 142/2, project C02 and via project SH 1228/3-1.  We also thank the PC$^2$ (Paderborn Center for Parallel Computing) for providing computing time.

\bibliography{database}


%\bibliography{your_external_BibTeX_file}

%=====================================
% References, variant B: internal bibliography
%=====================================

\end{document}